\documentclass[12pt]{iopart}

\usepackage{iopams}
\usepackage[utf8]{inputenc}
\usepackage[usenames, dvipsnames]{color}
\usepackage{hyperref}
\usepackage{cite}
\begin{document}

\title[Resonances with point potentials]{Semitransparent One Dimensional Potential, A Green's Function Approach}

\author{F.H.~Maldonado-Villamizar}

\address{Centro de Investigación y de Estudios Avanzados del Instituto Politécnico Nacional, México, D.F.}
\ead{fmaldonado@fis.cinvestav.mx}
\begin{abstract}
We study the unstable harmonic oscillator and the unstable linear potential in the presence of the point potential, which is the superposition of the Dirac $\delta(x)$ and the derivative $\delta'(x)$. Using the \textit{physical} boundary conditions for the Green's function we derive for both systems the resonance poles and the resonance wave functions. The matching conditions for the resonance wave functions coincide with those obtained by the self-adjoint extensions of the point potentials and also by the modelling of the $\delta'(x)$. We find that, with our definitions, the pure $b\delta'(x)$ barrier is semi-transparent \textit{independent} of the value of $b$.
\end{abstract}


\section{Introducion}\label{sec:introducion}

The present paper is a contribution on the discussion about the physical properties of one dimensional point potentials as well as  the study of quantum models with resonances. As is well known,  point potentials are quantum perturbations of the Hamiltonian which are supported in one point or a countable number of them, either finite or denumerable. A typical example is the Dirac delta or its derivative. In the present work, our point potentials will be supported at the origin and will appear in combination with another potential term. 

An important property of point potentials is that they are often solvable or quasi-solvable. In the first case, we may find exact solutions for the eigenvalue problem associated to systems with this kind of potentials. In the second, energy values and other data, like resonance poles, can be obtained numerically as solutions of transcendental equations. This feature has its origin in the fact that point potentials are often semitransparent, i.e., they have non-vanishing transmission coefficient. Therefore, point potentials are useful tools to construct one-dimensional models with resonances, which will be of great help in the study of physical and mathematical properties of resonances.

The use of point potentials in quantum mechanics has a long tradition. It began with  Kronning-Penney~\cite{kron}, who used them as an approximation for a periodic medium. Next, Bethe and Peierls used a similar approximation for a nucleon model~\cite{bethe}.  Berezin and Faddev~\cite{fadeev} made a complete mathematical study of a three dimensional problem with such potentials. In addition, the study of point potentials has been also useful in other areas of physics for the solution of various types of problems, see for example~\cite{albeveiro}.

The construction of one-dimensional models with resonances making use of point potentials is not new.  Thus far, there are several studies on this field dealing with specific models \cite{zol1,zol2,zol3}. One of the most interesting  is the unstable quantum oscillator proposed by Espinosa and Kielanowski \cite{piotr}. It is so interesting not only because it shows all kind of features like bound and antibound states and resonances, but also because of the (sometimes unexpected) behavior of them under the variations of certain parameters \cite{gadella}. This one dimensional unstable quantum oscillator has a potential, which is a harmonic oscillator for $x<0$, zero for $x>0$ and at the origin a point potential of the form $a\delta(x)+b\delta'(x)$. Here, $a$ and $b$ are real parameters. Eventually, we can locate a discontinuity of the mass at the origin \cite{gadella}, which adds one more parameter (the ratio of masses) in the model description.

Although the characterization of a point potential of the Dirac  delta type, $\delta(x)$ is clear, this is usually not the case with its derivative $\delta'(x)$. The latter is a perturbation which has been defined in several non-equivalent ways \cite{seba,kurasov,GOL,ALB,BR}. In general, terms in the potential of the type $a\delta(x)+b\delta'(x)$ are defined through the theory of self adjoint extensions of symmetric operators with equal deficiency indices \cite{kurasov,RS}. This is also the case when dealing with self adjoint Hamiltonians in models with a discontinuity in the mass \cite{GKN}.

One equivalent method to study the properties of these models is through Green's functions associated to the Hamiltonians. As the Green's function can be looked as a kernel of the resolvent of the Hamiltonian, it is also a possible tool to study the resonances in the models under our consideration. Resolvents of self adjoint extensions of symmetric Hamiltonians can be obtained through the Krein formula, a technique which is not very familiar to physicists. Nevertheless, Green's function methods are much more familiar to the physicists. There are some precedents in the use of Green's functions with point potentials: for instance in the study of the variation of the energy levels of the infinite square well or the harmonic oscillator in presence of a point potential \cite{Larry}.

In this context, one of the obvious applications of Green's functions is the location of resonances (in terms of its energy and width) and other features like antibound states in one dimensional models with point potentials. In the present paper, we shall focus our attention in two models: the unstable harmonic oscillator and a similar linear potential with a perturbation of the type   $a\delta(x)+b\delta'(x)$. 

It is important to remark that a definition is usually needed for a self-adjoint determination of the Hamiltonian with the perturbation $a\delta(x)+b\delta'(x)$.  To this end, the  definition proposed in \cite{kurasov} is widely used, for instance in  \cite{Gadella1,GNN,gadella}. This definition is based in the use of matching conditions at the origin. 

However, we shall here adopt a different point of view. We first assume that both distributions the Dirac delta $\delta(x)$ and its derivative $\delta'(x)$ may be multiplied by functions having, along to their derivatives, a finite discontinuity at the origin. We have to define these type of products that also are distributions.  Here, instead of using the usual product proposed by Kurasov \cite{kurasov}, we prefer instead a more general form given by Zolotaryuk \cite{zol1}. The Schr\"odinger equation with the perturbation $a\delta(x)+b\delta'(x)$ becomes an equation between distributions. 

Then, the use of the Green's function, that in our case can be exactly evaluated, and of its properties, like their boundary conditions, will finally determine the matching conditions required for the self adjoint determination of the Hamiltonian with potential $a\delta(x)+b\delta'(x)$. Although this has been done for two particular cases, we think that the procedure is quite general. This is important, since it has been some confusion on the proper definition of the perturbation $a\delta(x)+b\delta'(x)$, due to the presence of the term on $\delta'(x)$ that in our approach is defined unambigously.

The plan of the paper is the following. In Section~\ref{sec:2} we fix our notation and include a general discussion of the Green's function of the Schrödinger equation, including the definition of the resonance.  In Section~\ref{sec:3} we discuss the Green's functions of two types of potential: harmonic oscillator and linear potential for $x<0$ and a free motion for $x>0$. These potentials do not have resonances and bound states, but the Green's functions for these potentials are used for the determination of the Green's functions of the point potentials in Section~\ref{sec:4}. In this section we also find conditions for resonances that follow from the Green's function. In Section~\ref{sec:5} we determine from the Green's functions the matching conditions for the wave functions of the resonances at the singularity (support) of the point potentials. Section~\ref{sec:conclusions} contains a discussion of our results and conclusions.

\section{The Green's Function  for the Schrödinger Equation}\label{sec:2}
The time independent Schrödinger Equation~(SE) is
\begin{equation}\label{scho8}
H\psi(x)=E\psi(x),
\end{equation}
where  $H$ is the Hamiltonian of the system,  $E$ is an energy eigenvalue of  $H$ and $\psi(x)$ is the associated eigenfunction.
A way to solve Eq.~(\ref{scho8}) is by using the solution of the equation
\begin{equation}\label{eq:6}
 ( H -z)G(x,x';z)=\delta(x-x').
\end{equation}
 $G(x,x';z)$ is called the Green's Function (GF) for the operator   $H$ and $z\in \mathbb{C}$, see for example \cite{economou}. In general  $G(x,x';z)$ is an analytic function in the whole plane  $z$, except for a set of points, which can be continuous or discrete. Such a set contains the bound states energies and the resonances values of the system and it is called the spectrum of~$H$. It is denoted by  $\mathrm{Sp}(H)$.

For $z\notin\mathrm{Sp}(H)$  the solution of the equation
\begin{equation}\label{eigen}
 (H-z)\psi(x)=\phi(x),
\end{equation}
can be written as
\begin{equation}
 \psi(x)=\int G(x,x';z)\phi(x')dx',
\end{equation}
whereas for  $z \in\mathrm{Sp}(H)$ the solution of Eq.~(\ref{eigen}) is
\begin{equation}\label{eq:9}
\fl 
\psi(x)=\int G(x,x';z^+)\phi(x')dx'+\psi_{0}(x),\quad \mbox{with}\quad z^{+}=\lim_{b\to 0^{+}} (a+ i b)\quad a,b \in\mathbb{R}.
\end{equation}
 The term $\psi_0(x)$ is the solution to the homogeneous equation
\begin{equation}\label{eigen1}
 (H-z)\psi_0(x)=0.
\end{equation}

The function  $G(x,x';z)$ usually has a branch cut that in general can be identified to the positive real semi-axis. In adition, it may have  poles in the real axis, which correspond to  bound states and poles in the analytic continuation of the variable $z$ of $G(x,x';z)$ (which in the language of Riemann surfaces are located on the second Riemann sheet) of the form

\begin{equation}
 z=E_{0}-i\frac{\Gamma}{2}.
\end{equation}
For $\Gamma > 0$  there is an state with a finite mean lifetime, given by:
\begin{equation}
\tau=\frac{1}{\Gamma}.
\end{equation}

Let  $G_0(x,x';z)$ be the GF for the Hamiltonian $H_0$ and let   $H=H_0+\hat{V}$ be the new Hamiltonian under our scrutiny. Using the GF $G_0(x,x';z)$ of $H_0$, then the GF for  $G(x,x'z)$ of  $H$ can be formally calculated. To do this one can write
 Eq.~(\ref{eq:6})  as
\begin{equation}\label{eq:15}
 \left(H_{0}-z\right)G(x,x';z)=\delta(x-x')-\hat{V} (x)G(x,x';z).
\end{equation}
As  $G_0(x,x';z)$ satisfies
\begin{equation}\label{gren0}
(H_0-z)G_0(x,x';z)=\delta(x-x'),
\end{equation}
Eq.~(\ref{eq:15}) can be multiplied  from the left by $G_{0}(y,x;z)$ and one obtains
\begin{equation}\label{formint}
\fl\hspace{1.5 cm}\delta(y-x)G(x,x';z)=G_0(y,x;z)\delta(x-x')-G_0(y,x;z)\hat{V}(x)G(x,x';z).
\end{equation}
Formally,  Eq.~(\ref{formint}) can be integrated and $G(x,x';z)$ can be written as
\begin{equation}\label{eq:15a}
G(y,x';z)=G_0(y,x';z)-\int G_0(y,x;z)\hat{V}(x)G(x,x';z)dx.
\end{equation}

This derivation is well know and presented here as purely formal for the sake of completeness of our presentation.

\section{The Green's Function for the Unstable Quadratic and  Linear Potentials }\label{sec:3}

Along this section we shall derive the Green's functions for two potentials as mentioned in the Introduction. The first one will be the so called unstable harmonic oscillator, which is a semi-oscillator with a point potential supported in the origin. The  second one will be similar. We call it the linear unstable potential, as the semi-oscillator is replaced by a term in the form of half straight line.

\subsection{The free GF for the unstable quadratic potential}\label{sec:3a}

Let us start with the {\it unstable harmonic oscillator}. It was introduced by by Espinosa and Kielanowski in their study of resonances, see~\cite{piotr}. The form of the potential is a harmonic oscillator for $x<0$ (and hence called semi-oscillator) plus a point potential supported at the origin and of the form  $a\delta(x)+b\delta'(x)$, $a$ and $b$ being real numbers. The original idea in \cite{piotr} was to introduce a Dirac delta in order to produce resonances in the model, although it was shown in \cite{Gadella1} that resonances appear even in the absence of any kind of point potential. In the mentioned paper \cite{Gadella1}, we also have made a thoroughly analysis of the behavior of resonances for this model.

Now, we begin our analysis by defining the {\it unperturbed} Hamiltonian $H_0$ as:
\begin{equation}\label{ham1}
H_0=-\frac{1}{2}\frac{d^2}{dx^2}+\frac{x^2}{2}\theta(-x).
\end{equation}

Later, we shall define the total Hamiltonian as the sum of $H_0$ plus the point potential. 
To obtain the GF $G_0(x,x';z)$, for $H_0$ in Eq.~(\ref{ham1}), it is necessary to take  Eq.~(\ref{gren0}) and observe that there exist two equations
\begin{equation}\label{gensres0}
\begin{array}{l}
(H_0-z)G_0^+(x,x';z)=0,\qquad\mbox{for}\qquad x>x';\\
(H_0-z)G_0^-(x,x';z)=0,\qquad\mbox{for}\qquad x<x'.
\end{array}
\end{equation}
For each of the equations in Eq.~(\ref{gensres0}), there are another two equations, one for  $x>0$ and the other for $x<0$. Using the set of equations~(\ref{gensres0}) along with the following conditions:
\begin{enumerate}
\item[\textit{i)}] continuity of $G_0^{\pm}(x,x';z)$
\begin{equation}\label{contg}
\lim_{\epsilon\to 0}\left(G_0^{\pm}(x,x+\epsilon;z)-G_0^{\pm}(x,x-\epsilon;z)\right)=0,
\end{equation}
\item[\textit{ii)}] jump discontinuity for the derivative with respect to the $x$ (denoted by the superscript $(1)$)
\begin{equation}\label{contdg}
\lim_{\epsilon\to 0}\left(G_0^{\pm(1)}(x,x+\epsilon;z)-G_0^{\pm(1)}(x,x-\epsilon;z)\right)=-2,
\end{equation}
\item[\textit{iii)}] purely outgoing boundary condition for $G_0^+(x,x';z)$ for $x\to\infty$
\begin{equation}\label{pobc1}
G_0^+(x,x';z)\sim \mbox{e}^{ikx}\qquad\mbox{for}\qquad x\to \infty,
\end{equation}
\item[\textit{iv)}] localization of the $G_0^-(x,x';z)$
\begin{equation}\label{pobc2}
G_0^-(x,x';z)\to 0 \qquad\mbox{for}\qquad x\to -\infty,
\end{equation}
\end{enumerate}
the GF $G_0(x,x';z)$ can be calculated explicitly. The conditions \textit{i)} and \textit{ii)} can be deduced from  Eq.~(\ref{gensres0}) and \textit{iii)} and \textit{iv)} are well established  physical requirements \cite{Newton}. 

Then, let us use  Eq. (\ref{gensres0}) with the Hamiltonian $H_0$ given in  Eq.(\ref{ham1}). We obtain that the GF satisfying the conditions given by  Eq.(\ref{contg}), Eq(\ref{contdg}), Eq.(\ref{pobc1}) and Eq.(\ref{pobc2}) is given by:
\begin{equation}\label{g+o}
 \fl G_0^+(x,x';\epsilon)=A(x')\left\{
 \begin{array}{lcc}y_1(2\epsilon,x)-\frac{i}{k}y_2(2\epsilon,x)&\mbox{    }& \mbox{for\;} x<0\\&&\\
\mbox{e}^{ikx}&&
  \mbox{for\;} x>0
 \end{array}\right.
\end{equation}
\begin{equation}\label{g-o}
 \fl G_0^-(x,x';\epsilon)=C(x')\left\{
 \begin{array}{lcc}y_1(2\epsilon,x)+2g(\epsilon)y_2(2\epsilon,x)&& \mbox{for\;} x<0\\&&\\
 \frac{1}{2}\left(1+\frac{2ig(\epsilon)}{k}\right)e^{ikx}+ \frac{1}{2}\left(1-\frac{2ig(\epsilon)}{k}\right)e^{-ikx}&&
  \mbox{for\;} x>0
 \end{array}\right.\,,
\end{equation}
with
\begin{equation}
\begin{array}{l}
\displaystyle A(x')=\frac{ \frac{1}{2}\left(1+\frac{2ig(\epsilon)}{\sqrt{z}}\right)e^{ikx'}+ \frac{1}{2}\left(1-\frac{2ig(\epsilon)}{\sqrt{z}}\right)e^{-ikx'}}{1-\frac{2g(\epsilon)}{ik}},\\
\displaystyle C(x')=\frac{\mbox{e}^{ikx'}}{1-\frac{2g(\epsilon)}{ik}},\\
\displaystyle g(\epsilon)=\Gamma\left(\frac{3-2\epsilon}{4}\right)\Gamma\left(\frac{1-2\epsilon}{4}\right)^{-1}.
\end{array}
\end{equation}

The functions $y_1(a,x)$ and $y_2(a,x)$ are the parabolic cylindric functions and $\Gamma(z)$ is the Gamma function (see for example~\cite{abramowitz}). It is a simple exercise to check that (19) and (20) satisfy all the properties required for the Green's functions.

\subsection{The free GF for the unstable linear potential}\label{sec:3b}

The case of the linear potential is quite interesting, since it is one of the few potentials for which the Schrödinger equation admits  a known analytic solution. In addition, it appears in the first order term in the \emph{WKB} method  for SE , see \cite{landau}. In this case, the free Hamiltonian $H_0$ is given by
\begin{equation}\label{linear}
H_0(x)=\frac{1}{2}\frac{d^2}{dx^2}-Fx\theta(-x),
\end{equation} 
where $F$ is a positive constant often interpreted as the intensity of a constant electric field \cite{ludv}.

Following the same procedure as the previous case and taking into account the mentioned properties that the GF must have, one finds the GF,  $G_0(x,x';z)$, as:%
\begin{equation}\label{g+l}
\fl G_0^+(\xi,\xi';\epsilon)=\frac{C^-_i(-\xi'-\epsilon)}{C_i^{+,}(-\epsilon)+ikC_i^+(-\epsilon)} \left\{\begin{array}{ll}
-i\mathrm{e}^{ik\xi}&\xi>0\\&\\
i\pi(\alpha A_i(-\xi-\epsilon)-\beta B_i(-\xi-\epsilon)) &\xi<0
\end{array}\right.
\end{equation}
\begin{equation}\label{g-l}
\fl G_0^{-}(\xi,\xi';\epsilon)=\frac{\pi(\alpha A_i(-\xi'-\epsilon)-\beta B_i(-\xi'-\epsilon))}{C_i^{+,}(-\epsilon)+ikC_i^+(-\epsilon) } \left\{\begin{array}{ll}
\frac{1}{2k}\left(\gamma\mathrm{e}^{ik\xi}+\sigma\mathrm{e}^{-ik\xi}\right)&\xi>0\\&\\
iC_i(-\xi-\epsilon) &\xi<0
\end{array}\right.\,,
\end{equation}
where
\begin{eqnarray}
\nonumber
\begin{array}{ccc}
x=\left(\frac{1}{2F}\right)^{1/3}\xi,&
\;\epsilon=2\left(\frac{1}{2F}\right)^{2/3}E,&
\; k^2=\epsilon
\end{array}\\
\fl
\begin{array}{ccc}
\alpha=B_i'(-\epsilon)+ikB_i(-\epsilon),&
\;\beta=A_i'(-\epsilon)+ikA_i(-\epsilon),&
\;\gamma=C_i'(-\epsilon)+ikC_i(-\epsilon).
\end{array}\\
\begin{array}{cc}\nonumber
\sigma=-C_i'(-\epsilon)+ikC_i(-\epsilon),&
\; C_i^{\pm}(x)=A_i(x)\pm iB_i(x).
\end{array}
\end{eqnarray}

The functions $A_i(x)$, $B_i(x)$ and $C_i(x)$ are the Airy functions. One can find their properties in~\cite{abramowitz}.

Once we have obtained the explicit forms for the GF of the free Hamiltonian for both cases under our study, it is time to add the perturbation on the form of point potential. This will be done in the next section.

\section{Point Potentials}\label{sec:4}

Next, we shall construct the Green's functions corresponding to the Hamiltonians studied in the previous section, perturbed with a point potential. In the present case, we have a total Hamiltonian of the form
\begin{equation}\label{fullH}
H=H_0+V(x),
\end{equation}
where  $H_0$ is either (13) or (22), for which the Green's function $G_0(x,x';z)$ has been already derived and the perturbation $V(x)$ is the contribution of the point potential given by
\begin{equation}\label{fullV}
V(x)=a\delta(x)+b\delta'(x)\,.
\end{equation}

In order to give a proper self-adjoint determination for (\ref{fullH}), we need the theory of self-adjoint extensions of symmetric operators with equal deficiency indices. This theory was originally proposed by von Neumann \cite{RS}. In order to choose the domain for the self-adjoint determination of (\ref{fullH}), we need to give matching conditions at the origin \cite{kurasov}. This allows to construct all the self adjoint determinations of our Hamiltonians \cite{GNN}. Note that the functions in the domain of $H$ show in general a discontinuity at the origin.  

To obtain the GF for $H$ as in (26), we insert (\ref{fullV}) into  (\ref{eq:15a}), so that: 
\begin{eqnarray}\nonumber
\hspace*{-20pt}G(x,x';z)&=& G_0(x,x',z)-a\int G_0(x,x'',z)\delta(x'')G(x'',x',z)dx''
 \\&&-b\int G_0(x,x'',z)\delta'(x'')G(x'',x',z)dx''\,,\label{eq:32}
\end{eqnarray}
where $G(x,x';z)$ is the GF of $H$ as in  (\ref{fullH}). This is an integral equation for  $G(x,x';z)$, which is solvable in our case due to the specific form of the potential (27). However, one should take into account that $G(x,x';z)$ and its partial derivatives may be discontinuous in its arguments $x$ and $x'$. Then, it is necessary to define the action of the Dirac delta $\delta(x)$ and its derivative $\delta'(x)$ on functions showing a finite jump at the origin. This includes the form of the products $f(x)\delta(x)$ and $f(x)\delta'(x)$, where $f(x)$ is an arbitrary function with discontinuity at the origin. A first definition has been proposed by Kurasov \cite{kurasov} and applied with similar purposes  in other papers \cite{gadella, Gadella1,GNN}. A more general version has been proposed by Zolotaryuk \cite{zol,zol1,zol2,zol3} and this will be used here. This allows to view the problem with a greater generality. Then, the formulas for the products $f(x)\delta(x)$ and $f(x)\delta'(x)$ we have chosen are defined by:
\begin{equation}\label{integral}
\begin{array}{l}
\displaystyle\int f(x)\delta(x)dx= \zeta f(0+)+\eta f(0-),\\[2ex]
\displaystyle\int f(x)\delta'(x)dx=-(\zeta f'(0+)+\eta f'(0-)).
\end{array}
\end{equation}
where $\zeta$ and $\eta$ are non-negative real numbers such that $\zeta+\eta=1$. For any function $f(x)$, we denote by $f(0+)$ and $f(0-)$ the right and left limits at the origin, respectively. 

From now on, we shall use the following abridged notation: $G(x,x')=G(x,x';z)$.  Then, using (\ref{integral}) in (\ref{eq:32}), we obtain the following equation:
\begin{eqnarray}\label{int1}\nonumber
 \hspace*{-40pt}G(x,x')&=& G_0(x,x')-a(\eta+\zeta) G_0(x,0)G(0,x')+b G_0(x,0)\left(\zeta G^{(1)}(0+,x')\right)\\
 && \label{eq33} +\eta G^{(1)}(0-,x')
 +b\left(\zeta G^{(2)}(x,0+)+\eta G^{(2)}(x,0-)\right)G(0,x').
\end{eqnarray}
The superscripts $(1)$ and $(2)$ in   $G^{(1)}(x,y)$ , $G^{(2)}(x,y)$ refer to the partial derivatives of $G(x,y)$ with respect to the first and second variable, respectively and  $G^{(12)}(x,y)$ is the mixed, second derivative. In (\ref{int1}), we have used the following notation
\begin{equation}\label{eq34}
G(0\pm,y)=\lim_{x\to 0\pm}G(x,y).
\end{equation}

Let us go back to~(\ref{eq33}). This is a relation between the GF of $H_0$ and the GF for $H$ and their partial derivatives. Fortunately, it can be solved so that $G(x,x')$ can be written in terms of $G_0(x,x')$ including the partial derivatives of both GF.  Needless to say that, in order to solve (\ref{int1}), the functions $G(0,x')$, $G^{(1)}(0\pm,x')$, $G^{(2)}(0\pm,x')$ and the value $G^{(12)}(0,0)$ have to be known. 

Then, if we take the derivative in~(\ref{eq33}) with respect to $x$, we obtain:
\begin{eqnarray}\nonumber
 G^{(1)}(x,x')&=&G_0^{(1)}(x,x')-a(\zeta+\eta) G_0^{(1)}(x,0)G(0,x')\\[2ex]
 &&\nonumber+bG_0^{(1)}(x,0)\left(\zeta G^{(1)}(0+,x')+\eta G^{(1)}(0-,x')\right)\\[2ex]
 &&+b\left(\zeta G^{(12)}(x,0+)+\eta G^{(12)}(x,0-)\right)G(0,x')\,,\label{eq33n}
\end{eqnarray}

Now, first take the limits $x\to 0^+$ and  $x\to 0^-$ in~(\ref{eq33n}), i.e., the limits in  when $x$ goes to zero to the right and to the left, respectively. One finds for $x\to 0^+$
\begin{eqnarray}\nonumber
 G^{(1)}(0+,x')&=&G_0^{(1)}(0+,x')-a(\zeta+\eta) G_0^{(1)}(0+,0)G(0,x')\\[2ex]
 &&\nonumber+bG_0^{(1)}(0+,0)\left(\zeta G^{(1)}(0+,x')+\eta G^{(1)}(0-,x')\right)\\[2ex]
 &&+b\left(\zeta G^{(12)}(0+,0+)+\eta G^{(12)}(0+,0-)\right)G(0,x')\,,
 \label{0+}
\end{eqnarray}
and for $x\to 0^-$
\begin{eqnarray}\nonumber
 G^{(1)}(0-,x')&=&G_0^{(1)}(0-,x')-a(\zeta+\eta) G_0^{(1)}(0-,0)G(0,x')\\[2ex]
 &&\nonumber+bG_0^{(1)}(0-,0)\left(\zeta G^{(1)}(0+,x')+\eta G^{(1)}(0-,x')\right)\\[2ex]
 &&+b\left(\zeta G^{(12)}(0-,0+)+\eta G^{(12)}(0-,0-)\right)G(0-,x')\,.\label{0-}
\end{eqnarray} 

Equations~(\ref{0+}) and~(\ref{0-}) look rather complicated. Nevertheless, they can be simplified and written in a compact form as the following simple system of equations: 
\begin{equation}\label{lineq}
\begin{array}{l}
\displaystyle AG^{(1)}(0+,x')+BG^{(1)}(0-,x')=G^{(1)}_0(0+,x')+CG(0,x')\,,\\[2ex]
\displaystyle DG^{(1)}(0+,x')+EG^{(1)}(0-,x')=G^{(1)}_0(0+,x')+FG(0,x')\,.
\end{array}
\end{equation}
In~(\ref{lineq}), we have made use of the following definitions:
\begin{equation}
\begin{array}{l}
A=\left(1-b\zeta G_0^{(1)}(0+,0)\right),\quad
B=-b\eta G_0^{(1)}(0+,0),\\
C=(\zeta+\eta)\left(bG^{(12)}(0+,0-)-a G_0^{(1)}(0+,0)\right),\\
D=-b\zeta G_0^{(1)}(0-,0),\quad
E=\left(1-b\eta G_0^{(1)}(0-,0)\right),\\
F=(\zeta+\eta)\left(bG_0^{(12)}(0-,0+)-a G_0^{(1)}(0-,0)\right),\\
\Delta=1-b\left(\zeta G_0^{(1)}(0+,0)+\eta G_0^{(1)}(0-,0)\right).
\end{array}
\end{equation}
Now, we can solve system (\ref{lineq}), in terms of $G^{(1)}(0+,x')$ and $G^{(1)}(0+,x')$. This gives:
\begin{equation}
\hspace*{-30pt}\begin{array}{l}
\displaystyle G^{(1)}(0+,x')=\frac{EG^{(1)}_0(0+,x')-BG^{(1)}_0(0-,x')+(EC-BF)G(0,x')}{\Delta}\,,\\[2ex]
\displaystyle G^{(1)}(0-,x')=\frac{AG^{(1)}_0(0-,x')-DG^{(1)}_0(0+,x')+(AF-DC)G(0,x')}{\Delta}\,.
\end{array}
\end{equation}

To solve(\ref{eq33}), we still need the factor $G(0,x')$. This can be found by taking the limit  $x\to 0$ in~(\ref{eq33}), which gives
\begin{eqnarray}\nonumber\label{g0}
G(0,x')&=& G_0(0,x')-a(\zeta+\eta) G_0(0,0)G(0,x')\\[2ex]
&&\nonumber +bG_0(0,0)\left(\zeta G^{(1)}(0+,x')+\eta G^{(1)}(0-,x')\right)\\[2ex]
&&+b(\zeta G_0^{(2)}(0,0+)+\eta G_0^{(2)}(0,0-))G(0,x')\,.
\end{eqnarray}

In order to avoid unpleasant long formulas, we simplify the notation. In fact, we write (39) as
\begin{equation}
 G(0,x')=\frac{1}{\lambda}G_0(0,x')+\frac{\lambda_1}{\lambda}G_0^{(1)}(0-,x')+\frac{\lambda_2}{\lambda}G_0^{(1)}(0+,x')\,,
\end{equation}
with
\begin{eqnarray}\nonumber
 \lambda&=& 1-\left(b(\zeta G^{(2)}_0(0,0+)+\eta G_0^{(2)}(0,0-))-a(\zeta+\eta)G_0(0,0)\right)\\[2ex]
&&-b\frac{G_0(0,0)}{\Delta}(\eta(AF-DC)+\zeta(EC-BF)),\\[2ex]
\lambda_1&=&b \frac{G_0(0,0)}{\Delta}(\eta A-\zeta B),\quad
\lambda_2=b\frac{G_0(0,0)}{\Delta}(\zeta E-\eta D)\,.
\end{eqnarray}

Finally, we carry all these results in~(\ref{eq33}), so as to write the 
 GF $G(x,x')$ in terms of $G_0(x,x')$ and its derivatives as\pagebreak
\begin{eqnarray}\nonumber
\hspace*{-40pt}G(x,x')&=& G_0(x,x')-\frac{\alpha_1}{\lambda}G_0(x,0)G_0(0,x')+\frac{\alpha_2}{\lambda}G_0(x,0)G_0^{(1)}(0-,x')\\[2ex]
&&\nonumber+\frac{\alpha_3}{\lambda}G_0(x,0)G_0^{(1)}(0+,x')+\frac{b\zeta}{\lambda}G_0^{(2)}(x,0+)G_0(0,x')\\[2ex]
&&\nonumber+
\frac{b\zeta\lambda_1}{\lambda}G_0^{(2)}(x,0+)G^{(1)}_0(0-,x')+\frac{b\zeta \lambda_1}{\lambda}G_0^{(2)}(x,0+)G^{(1)}_0(0+,x')\\[2ex]
&&\nonumber+
\frac{b\eta}{\lambda}G_0^{(2)}(x,0-)G_0(0,x')+\frac{b\eta\lambda_1}{\lambda}G_0^{(2)}(x,0-)G^{(1)}_0(0-,x')\\[2ex]
&&+\frac{b\eta \lambda_2}{\lambda}G_0^{(2)}(x,0-)G^{(1)}_0(0+,x')\,.
\end{eqnarray}
Here,
\begin{eqnarray*}
\fl \hspace*{30pt}\alpha_1=\frac{1}{\Delta}\left(b(\zeta(EC-BF)+\eta(AF-DC)-a\Delta (\zeta+\eta)\right)\,,\\[2ex]
\fl \hspace*{30pt}\alpha_2=\frac{1}{\Delta}\left(b(\zeta(-B\lambda +(EC-BF)\lambda_1)+\eta(A\lambda +(AF-DC)\lambda_1)-a\Delta \lambda_1(\zeta+\eta)\right)\,,\\[2ex]
\fl \hspace*{30pt}\alpha_3=\frac{1}{\Delta}\left(b(\zeta(E\lambda -(EC-BF)\lambda_2)-\eta(D\lambda -(AF-DC)\lambda_2)-a\Delta \lambda_2(\zeta+\eta)\right)\,.\\
\end{eqnarray*}

As we have already mentioned, the GF contains a lot of information about the system under consideration. In particular, isolated singularities of the GF $G(x,x';z)$ in the variable $z$ give not only energy values but also complex singularities provide energy and mean life of resonances.   Thus, in order to find the location of resonance poles, we choose $\lambda=0$ in (43) and take into account that resonance poles must have non-zero imaginary part, so that
\begin{eqnarray}\label{root}\nonumber
 1-&&\left(b(\zeta G^{(2)}_0(0,0+;z)+\eta G_0^{(2)}(0,0-;z))-a(\zeta+\eta)G_0(0,0;z)\right)\\
&&-b\frac{G_0(0,0;z)}{\Delta}(\eta(AF-DC)+\zeta(EC-BF))=0\,.
\label{eq:52}
\end{eqnarray}

Then, after some simplifications in (\ref{root}) and taking into account the properties of GF,  we arrive to:
\begin{equation}
 1-b( G^{(1)}_0(0,0+;z)+ G_0^{(1)}(0,0-;z))+4b^2\eta\zeta+aG_0(0,0;z)=0\,.\label{eq:52a}
\end{equation}

All these formulas can be applied no matter which is the chosen free Hamiltonian~$H_0$, provided that the potential be given by (\ref{fullV}). This obviously includes both situations under our consideration. For the case of the unstable harmonic oscillator, one obtains from (\ref{eq:52a}), along with (\ref{g+o}) and~(\ref{g-o}) the following equation given for resonance poles:
\begin{equation}\label{eq:53}
ik=\frac{2a}{(1+2b+4\zeta\eta b^2)}+2\frac{(1-2b+4\zeta\eta b^2)}{(1+2b+4\eta\zeta b^2)}\, g(\epsilon)\,.
\end{equation}
from which one can obtain the resonance energy and width. Note that the resonance condition depends on the parameters $\zeta$ and $\eta$, but in Eq.~(\ref{eq:53}) only one variable is included, because they satisfy the relation $\zeta+\eta=1$.

For the case  of the linear potential from Subsection~\ref{sec:3b} with point potential added, using again the formula~(\ref{eq:52a}) and replacing the values for Eqs.~(\ref{g+l}) and (\ref{g-l}). One obtains 
\begin{equation}\label{resairy}
ik=\frac{2a}{(1+2b+4\zeta\eta b^2)}+\frac{C_i^{,-}(-\epsilon)}{C_i^-(\epsilon)}\frac{(1-2b+4\zeta\eta b^2)}{(1+2b+4\eta\zeta b^2)}.
\end{equation}

One can note that Eqs.~(\ref{eq:53}) and~(\ref{resairy}) have the same structure.  This suggests the following:  Let $H_0$ be a free hamiltonian for which the exact solution is known and let  $V_0(x)$ be a continuous potential  with the following behavior:
\begin{equation}\label{vpro}
V_0(0)=0\qquad \mathrm{and}\qquad V_0(x)\to \infty \qquad \mathrm{as}\qquad x\to -\infty\,.
\end{equation}

Then, the resonance condition should have the structure implied by Eqs.~(\ref{eq:53}) and~(\ref{resairy}). This indeed is the case, because
if $G_0(x,x';z)$ is the Green's function for $H_0$, then from  Eq.~(\ref{eq:52a}) the general resonance condition can be calculated. If  $G^-_0(x,0;E)$ is denoted by $u(x,0;E)$,  the the resonance condition is then,
\begin{equation}
ik=\frac{u'(0,E)}{u(0,E)}\frac{(1-2b+4\zeta\eta b^2)}{(1+2b+4\zeta\eta b^2)}+\frac{2a}{(1+2b+4\zeta\eta b^2)}\,.
\end{equation}

Then, we conclude that if we can obtain the function $u(0,E)$, which is position independent, then we should be able to obtain the complex values of $E$  which are the resonance energies for the Hamiltonian pair $\{H_0,H=H_0+V\}$.

\section{The matching conditions}\label{sec:5}

\subsection{The matching conditions using the Green's Function}\label{sec:5a}

As we have mentioned earlier, Hamiltonians with point potentials are self adjoint extensions of symmetric Hamiltonian with equal deficiency indices. These extensions are determined by choosing their domains, i.e., the space of (square integrable) functions on which the extension acts. The functions on these domains are, in general discontinuous at the points supporting the potentials and they are determined by defining some  matching conditions at these points, which, in our case, it is only one: the origin. The forms of domains and, therefore, the form of  matching conditions give the conditions for the self-adjointness of the extension.   

In general, we give {\it a priori} the matching conditions that have to fulfill at the origin the functions in the domain of $H=H_0+a\delta(x)+b\delta'(x)$. In our case, we have not chosen these matching conditions prior to our discussion. In fact, it is not important, since these matching conditions appear as a result of our previous discussion.  

In order to see it, let us consider the Schrödinger equation for the harmonic semi-oscillator with the point potential $a\delta(x)+b\delta'(x)$, which reads 
\begin{equation}\label{eq:1'}
-\frac{1}{2}\frac{d^2}{dx^2}\psi(x)+\frac{x^2}{2}\psi(x)\theta(x)-E\psi(x)=-a\delta(x)\psi(x)-b\delta'(x)\psi(x).
\end{equation}

The objective is to obtain the form of the discontinuity of the solutions of~(\ref{eq:1'}), $\psi(x)$, and of their first derivatives, $\psi'(x)$, at the origin. 
Now, using Eq.~(\ref{eq:9}) and the GF $G_0(x,x')$, from Eq.~(\ref{gren0}), one has
\begin{equation}\label{eq:58}
\hspace*{-30pt}\psi(x)=\psi_0(x)-a\int G_0(x,x')\delta(x')\psi(x')dx'-b\int G_0(x,x')\delta'(x')\psi(x')dx'\,,
\end{equation}
where $\psi_0(x)$ is the solution of the equation without the $a\delta(x)+b\delta'(x)$ perturbation. 

Next, integrating Eq.~(\ref{eq:58}) and using relations~(\ref{integral}), we obtain the following equation:
\begin{eqnarray}\nonumber
\fl\psi(x)=\psi_0(x)-aG_0(x,0)\left(\zeta\psi(0+)+\eta\psi(0-)\right)
+b\left( G_0(x,0)(\zeta\psi'(0+)+\eta\psi'(0-)\right)\\[2ex]
+b(\zeta G_0^{(2)}(x,0+)\psi(0+)+\eta G_0^{(2)}(x,0-)\psi(0-))\,.\label{eq:59}
\end{eqnarray}
Since the potential  $V_0(x)$ in $H_0$ is continuous at   $x=0$, we conclude that both  $\psi_0(x)$ and $\psi'_0(x)$ are continuous at origin. From Eq.~(\ref{gren0}) one can verify that $G_0(x,0)$ and $G_0^{(12)}(x,0)$ are continuous at  $x=0$, while  
\begin{equation}\label{eq:75}
G_0^{(1)}(0+,0)-G_0^{(1)}(0-,0)=-2,\quad\quad G_0^{(2)}(0,0+)-G_0^{(2)}(0,0-)=2.
\end{equation}

Using  Eq.~(\ref{eq:75}) in Eq.~(\ref{eq:59}) one gets the following expression for the discontinuity of $\psi(x)$ at the origin:
\begin{equation}\label{eq:60}
\fl \psi(0+)-\psi(0-)=2b(\zeta \psi(0+)+\eta \psi(0-) )\Rightarrow 
 \psi(0+)(1-2b\zeta)=\psi(0-)(1+2b\eta)\,.
\end{equation}

Then, take the derivative of~(\ref{eq:59}) with respect to $x$. We obtain:
\begin{eqnarray}\label{eq:61}\nonumber
\fl \psi'(x)=-aG_0^{(1)}(x,0)\left(\zeta\psi(0+)+\eta\psi(0-)\right)
+b\left( G_0^{(1)}(x,0)(\zeta \psi'(0+)+\eta \psi'(0-)\right)\\[2ex]
+b\left(G_0^{(12)}(x,0+)\zeta\psi(0+)+G_0^{(12)}(x,0-)\eta\psi(0-) \right).
\end{eqnarray}

 Using Eq.~(\ref{eq:75}) in Eq.~(\ref{eq:61}), we obtain the discontinuity of the derivative at the origin:
\begin{equation}\label{eq:62}
\fl \hspace*{50pt} \psi'(0+)-\psi'(0-)=2a\left(\zeta\psi(0+)+\eta\psi(0-)\right)
-2b\left( \zeta\psi'(0+)+\eta\psi'(0-)\right)\,,
\end{equation}
which can be transformed into
\begin{equation}\label{eq:63}
 \psi'(0+)\left(1+2b\zeta\right)-2a\zeta\psi(0+)=2a\eta\psi(0-)
+\left(1-2b\eta\right)\psi'(0-)\,.
\end{equation}

Equations~(\ref{eq:60}) and Eq.~(\ref{eq:63}) can be written together in a matrix form as,
\begin{equation}\label{eq:76}
\fl \left(\begin{array}{cc}
(1-2b\zeta)&0\\
-2a\zeta&(1+2b\zeta)
\end{array}\right)\left(\begin{array}{c}
\psi(0+)\\\psi'(0+)
\end{array}\right)=\left(\begin{array}{cc}
(1+2b\eta)&0\\
2a\eta&(1-2b\eta)
\end{array}\right)\left(\begin{array}{c}
\psi(0-)\\\psi'(0-)
\end{array}\right)\,,
\end{equation}
or equivalently,
\begin{equation}\label{match}
 \hspace*{-40pt}\left(\begin{array}{c}
\displaystyle\psi(0+)
\vphantom{\frac{(1-2b\zeta)(1-2b\eta)}{(1-2b\zeta)(1+2b\zeta)}}\\
\displaystyle\psi'(0+)
\vphantom{\frac{(1-2b\zeta)(1-2b\eta)}{(1-2b\zeta)(1+2b\zeta)}}
\end{array}\right)=\left(\begin{array}{cc}
\displaystyle\frac{1+2b\eta}{1-2b\zeta}&0\\
\displaystyle\frac{2a}{(1-2b\zeta)(1+2b\zeta)}&
\displaystyle\frac{1-2b\eta}{1+2b\zeta}
\end{array}\right)\left(\begin{array}{c}
\displaystyle\psi(0-)
\vphantom{\frac{(1-2b\zeta)(1-2b\eta)}{(1-2b\zeta)(1+2b\zeta)}}\\
\displaystyle\psi'(0-)
\vphantom{\frac{(1-2b\zeta)(1-2b\eta)}{(1-2b\zeta)(1+2b\zeta)}}
\end{array}\right).
\end{equation}
Eq.~(\ref{match}) is the final form of the matching conditions for   $\psi(x)$ and $\psi'(x)$ at the support of the point potentials, for $b\neq \pm 1/(2\zeta)$. Eqs.~(\ref{eq:60}), (\ref{eq:61}) and (\ref{eq:62}) for the cases  $\zeta=\eta=1/2$ and $b=\pm 1$ give the relations:
\begin{equation}\label{eq:70}
\begin{array}{l}
\displaystyle\psi(0\pm)=0,\\
\displaystyle\psi'(0\mp)=\mp\frac{a}{2}\psi(0\mp).
\end{array}
\end{equation}

The same method can be used to derive the matching conditions for the unstable linear potential and the result coincides with the one in Eq.~(\ref{match}).

\subsection{Matching conditions using the self-adjoint extension method}\label{sec:5b}
The models studied here in Section~\ref{sec:3} are both described the Hamiltonian of the following type
\begin{equation}\label{eq:1}
 H=H_0+V_0(x)+a\delta(x)+b\delta'(x)\,,
\end{equation}
where
\begin{equation}
H_0=-\frac{1}{2}\frac{d^2}{dx^2}\,.
\end{equation}
$V_0(x)$ satisfies the conditions given by Eq.~(\ref{vpro}). 

The two last terms in  Eq.~(\ref{eq:1}) can also obtained by using the self-adjoint extension method for the free Hamiltonian  $H_0$~\cite{kurasov}. In our paper the Hamiltonian without the point potential contains an additional self-adjoint term  $V_0(x)\theta(-x)$. Such a term does not alter the procedure used in~\cite{kurasov}, so that the Hamiltonian~(\ref{eq:1}) can be considered as the self-adjoint extension of the Hamiltonian $H=H_{0}+V_{0}(x)$. From this follows that that the wave function $\psi(x)$ and its derivative should obey the same matching conditionas at $x=0$ as in~\cite{kurasov}, which is~\footnote{Only the symmetric case $\zeta=\eta=1/2$ is considered in Ref.~\cite{kurasov}.}
\begin{equation}\label{eq:2}
\left(
\begin{array}{c} 
\displaystyle\psi(0+)\vphantom{\frac{(1-2b\zeta)(1-2b\eta)}{(1-2b\zeta)(1+2b\zeta)}}\\
\displaystyle\psi'(0+)\vphantom{\frac{(1-2b\zeta)(1-2b\eta)}{(1-2b\zeta)(1+2b\zeta)}}\end{array}
\right)=
\left(
\begin{array}{cc}
\displaystyle\frac{1+b}{1-b}&\displaystyle 0\\
\displaystyle\frac{2a}{1-b^2}&\displaystyle \frac{1-b}{1+b}
\end{array}\right)
\left(
\begin{array}{c}
\displaystyle \psi(0-)\vphantom{\frac{(1-2b\zeta)(1-2b\eta)}{(1-2b\zeta)(1+2b\zeta)}}\\
\displaystyle \psi'(0-)\vphantom{\frac{(1-2b\zeta)(1-2b\eta)}{(1-2b\zeta)(1+2b\zeta)}}
\end{array}\right),
\end{equation}
One can see that Eq.~(\ref{eq:2}) is not valid for the cases  $b=\pm 1$, the matching conditions for these values of  $b$ are given in~\cite{kurasov}. One can verify that Eq.~(\ref{eq:2}) is the same as in   Eq.~(\ref{match}) for the special case  $\zeta=\eta=1/2$ and the for cases $b=\pm 1$ the results of~\cite{kurasov} also coincide with   Eq.~(\ref{eq:70}).

\section{Conclusions}\label{sec:conclusions}
We have considered the unstable harmonic oscillator and unstable linear potential with point potentials. In both cases, we found that the resonances are formed and the point potential is semi transparent. Our method of derivation of the resonance condition and the matching condition for the wave function at the support of the point potential is based on the analysis of the Green's function of the system. From the poles positions of the Green's function we obtain the resonance conditions and from the residues of the poles we get the wave functions of the resonances. Our assumptions are minimal: we only assume the standard boundary conditions imposed by the \textit{physical requirenments}. We make no assumptions about the model of the $\delta'(x)$ as in Ref.~\cite{zol} or about the self-adjoint extensions for point interactions~\cite{kurasov}. It is remarkable, that our results for the matching condition of the wave function coincide with those of Refs.~\cite{zol,kurasov}.

It is also very interesting to note that within our approach the barrier formed by the pure $b\delta'(x)$ point potential is semi transparent for all values of the parameter $b\neq\pm1/(2\zeta)$. This is in contrast to the results from Ref.~\cite{zol}, where it was found that such a potential is semi-transparent only for certain discrete values of $b$.

This work shows the equivalence between the method of self-adjoint extensions for the Schrödinger operator with the method of Green's functions. Whereas the continuity conditions in the former are used only in one dimension the latter has the advantage, that it can be extended to several dimensions in a relatively easy way. One can also see the utility of the non-perturbative solution for the complete Green's function, which allows to avoid lengthy task of the perturbative calculation for the case of the singular potential with a delta function derivative.

\ack
The author would like to thank P. Kielanowski and M. Gadella for useful discussions and to  CONACyT for the support, under grant  No. 266294.

\end{document}